\documentclass[twocolumn, prl, showpacs, superscriptaddress]{revtex4}
\pdfoutput=1

\usepackage{graphicx}
\usepackage{amssymb}
\usepackage{amsmath}
{\catcode`\|=\active
\gdef\Braket#1{\left<\mathcode`\|"8000\let|\BraVert {#1}\right>}}
\def\BraVert{\egroup\,\vrule\,\bgroup}
\newcommand{\Ket}[1]{\ensuremath{\left\lvert{#1}\right\rangle}}

\newcommand{\ket}[1]{\ensuremath{\lvert{#1}\rangle}}

\usepackage{hyperref}
\hypersetup{colorlinks=true,linkcolor=black, citecolor=black, urlcolor=blue}

\graphicspath{{figures/}{figures/converted/}}

\newcommand{\topp}[1]{\ensuremath{^{({#1})}}}
\DeclareMathOperator{\sinc}{sinc}

\renewcommand{\vec}[1]{\mathbf{{#1}}}
\newcommand{\upst}{1}
\newcommand{\downst}{0}

\begin{document}
\title{Quantum computing with an electron spin ensemble}

\author{J.H.~Wesenberg} \affiliation{Department of Materials, University of Oxford, Oxford  OX1 3PH,  United Kingdom}
\author{A.~Ardavan} \affiliation{Clarendon Laboratory, Department of Physics, University of Oxford, Oxford OX1 3PH,  United Kingdom}
\author{G.A.D.~Briggs} \affiliation{Department of Materials, University of Oxford, Oxford OX1 3PH,  United Kingdom}
\author{J.J.L.~Morton} \affiliation{Department of Materials, University of Oxford, Oxford OX1 3PH,  United Kingdom}\affiliation{Clarendon Laboratory, Department of Physics, University of Oxford, Oxford OX1 3PH,  United Kingdom}
\author{R.J.~Schoelkopf} \affiliation{Department of Applied Physics, Yale University, New Haven, Connecticut 06520}
\author{D.I.~Schuster} \affiliation{Department of Applied Physics, Yale University, New Haven, Connecticut 06520}
\author{K.~M{\o}lmer} \affiliation{Lundbeck Foundation Theoretical Center for Quantum System Research, Department of Physics and Astronomy, University of Aarhus, 8000 Aarhus C, Denmark}

\date{\today}

\begin{abstract}
  We propose to encode a register of quantum bits in different
  collective electron spin wave excitations in a solid
  medium. Coupling to spins is enabled by locating them in
  the vicinity of a superconducting transmission line cavity, and
  making use of their strong collective coupling to the quantized
  radiation field. The transformation between different spin waves is
  achieved by applying gradient magnetic fields across the sample,
  while a Cooper Pair Box, resonant with the cavity field, may be used
  to carry out one- and two-qubit gate operations.
\end{abstract}

\maketitle

The construction of a large quantum computer is a challenge for
current research. The overarching problem is to develop physical
systems which can reliably store thousands of qubits and which allow
addressability of individual bits and pairs of bits in gate
operations. Proposals in which single trapped ions or atoms encode
qubits in their internal state have successfully demonstrated the
building blocks for few-bit devices, while scaling of these systems to
larger register sizes is believed to require interconnects, e.g.~with
optical transmission.  A novel collective encoding scheme for qubits
proposes to use many identical quantum systems to encode each qubit,
either in the collective population of different internal states
\cite{brion07:quantum,saffman08:scaling,tordrup08:quantum} or in
different spatial modes of excitation of the entire system
\cite{tordrup08:holographic,surmacz08:efficient}.

In this Letter we propose a hybrid approach to quantum computing making
use of an ensemble of $10^{10} - 10^{12}$ electron spins coupled to a
superconducting transmission line cavity.
We will describe how a large number of spatial modes can be addressed
in the spin ensemble, and how a transmon Cooper Pair Box (CPB)
\cite{koch07:charge-insensitive} integrated in the cavity can provide
one- and two-bit gates for quantum computing in the spin ensemble
\cite{wallraff04:strong,schoelkopf08:wiring}.  Our scheme enables
materials for which large spin coherence times have been demonstrated
in ensemble measurements to be incorporated into a solid state
device. In this way, without requiring single spin measurement or
strong coupling to a cavity, full use can be made of the sophisticated
techniques which are now well established for control of large numbers
of spins.

\begin{figure}
  \centering
  \includegraphics[width=\linewidth]{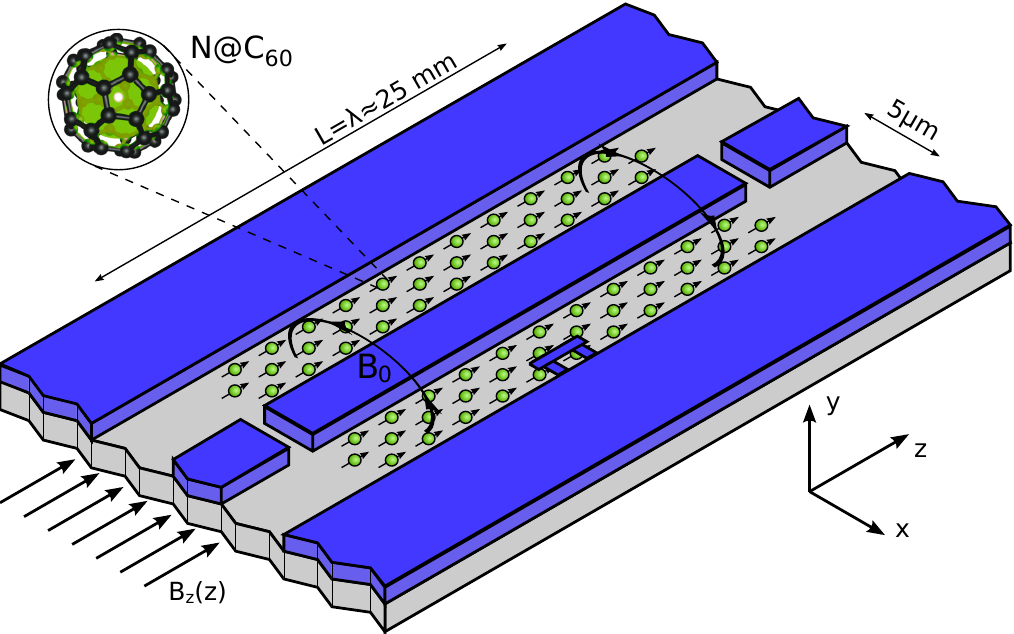}
  \caption{ \label{fig:setup} (Color online) 
    Physical setup, consisting of a superconducting transmission line
    cavity coupled to an ensemble of electron spins and a transmon Cooper
    Pair Box \cite{koch07:charge-insensitive}.  The cavity dimensions
    allow on the order of $N=10^{11}$ electron spins to be coupled to the
    cavity mode with an average coupling strength of $\bar{g}\approx
    2\pi\times 20 \,\text{Hz}$.
    An external magnetic field composed of a homogeneous bias field $B
    \hat{z}$ and a switchable linear gradient $( z \hat{z} - y
    \hat{y}) \Delta B/L$ is applied to the system.  }
\end{figure}

%
The proposed physical setup, as illustrated in Fig.~\ref{fig:setup},
consists of a superconducting transmission line cavity coupled to a
large number of solid-state electron spins doped into or deposited on
the surface of the substrate.
Two interesting choices for the electron spins would be P-doped Si and
endohedral fullerene molecules, e.g.~N@C$_{60}$, which would offer
spin coherence times up to tens of milliseconds
\cite{tyryshkin03:electron,
morton06:electron,
morton07:environmental}.
Hahn-echo techniques may be applied to counter inhomogeneous
broadening mechanisms, and the coherence time scale may even be
further extended by transferring the electron spin state to nuclear
spin degrees of freedom where coherence times exceeding seconds have
been demonstrated \cite{morton08:solid-state}.
The spins are biased with a homogeneous magnetic field $B$ in the
plane of the cavity, causing Larmor precession at an angular frequency
of $\omega_s = m_0 B/\hbar$, where $m_0$ is the magnetic dipole moment
of the spins. With a cavity resonance frequency of $\omega_c\approx
2\pi\times 5\,\text{GHz}$ a bias field of $B=180\,\text{mT}$ is
required to bring the spin precession into resonance. Even in the
presence of the bias field, cavity linewidths as low as $\kappa\approx
2\pi \times 250 \,\text{kHz}$ are possible \cite{yale:largebok}.

The $q^\text{th}$ spin located at $\vec{r}_q$ will couple to the cavity
annihilation and creation operators, $\hat{a}$ and $\hat{a}^\dagger$,
with a strength $g_q\equiv m_0 B_0(\vec{r}_q)/2 \hbar$ where
$B_0(\vec{r})$ is the zero-point magnetic field of the cavity mode. We
will use $\bar{g}\equiv\sqrt{\sum_q |g(\vec{r}_q)|^2/N}$ to denote the
average coupling strength.

We now consider the collective coupling between the spin ensemble and
the cavity.
Assuming the spin-cavity detuning $\Delta_s\equiv\omega_s-\omega_c$ to
be small compared to $\omega_{s,c}$, we can apply the rotating wave
approximation and describe the coupling by the interaction Hamiltonian
\begin{align}\label{HamCS}
  \hat{H}_\text{sc}&
  =\sum_{q=1}^N \hbar g_q \left( \hat{a}^\dagger \hat{\sigma}_-\topp{q}+\hat{a} \hat{\sigma}_+\topp{q} \right)
  -\hbar\Delta_s \hat{a}^\dagger \hat{a} \nonumber \\
  &=\hbar\, \sqrt{N} \bar{g}\left( \hat{a}^\dagger  \hat{b}+\hat{a}\hat{b}^\dagger\right)
  - \hbar \Delta_s \hat{a}^\dagger \hat{a},
\end{align}
where $\hat{\sigma}_-\topp{q}$ ($\hat{\sigma}_+\topp{q}$) is the Pauli
spin lowering (raising) operator of the $q^{th}$ spin, and we
introduce the collective spin lowering operator $\hat{b} \equiv
\tfrac{1}{\sqrt{N}} \sum_{q=1}^N \hat{\sigma}^{(q)}_- g_q/\bar{g}$ and
its Hermitian adjoint $\hat{b}^\dagger$ in the second line.
Assuming that the sample is strongly polarized, the collective
lowering and raising operators obey the commutator relation
$[\hat{b},\hat{b}^\dagger]=1$, so that the collective excitation behaves like a
harmonic oscillator degree of freedom. The lowest two spin oscillator
states are the state with all spins pointing down,
$\ket{\text{vac}}\equiv \ket{\downst}^{\otimes N}$, and the state with a
single collective excitation
\begin{equation} \label{psi1u}
\ket{\psi_1(0)} \equiv \hat{b}^\dagger \ket{\text{vac}} = \tfrac{1}{\sqrt{N}} \sum_q \frac{g_q}{\bar{g}} \Ket{\downst_1 \ldots \upst_{q} \ldots. \downst_N}.
\end{equation}
According to the second line of Eq.~(\ref{HamCS}) the cavity field and
the collective spin oscillator behave as two coupled oscillators with a
coupling strength that is enhanced with a factor of $\sqrt{N}$ relative
to the single spin coupling strength, so that any states of the two
systems are interchangeably mapped between them with the collective Rabi
frequency $\sqrt{N}\bar{g}$.
For the parameters given in Fig.~\ref{fig:setup}, the effective
coupling $\sqrt{N} \bar{g}$ is $\sim 2\pi \times \,6\,\text{MHz}$, and, thus,
exceeds by orders of magnitude the decay rates of collective spin
excitations (governed by the single spin decay rate
\cite{chase08:collective}) as well as the cavity decay rate $\kappa$.

The long coherence lifetimes in trapped atomic and molecular systems
have inspired recent proposals to transfer and store the cavity field
excitation in a single rotationally excited polar molecule
\cite{andre06:coherent}, and, to benefit from the collectively
enhanced coupling, in collectively excited states of many molecules
\cite{rabl06:hybrid}. In these proposals, molecules
would have to be cooled and trapped in the close vicinity of the cavity
transmission line, while the electron spins in the present proposal are being
held in a host material - at the price of shorter achievable coherence
times.
In a physical set-up similar to ours \cite{imamoglu08:cavity-qed}, the energy
splitting of the two lower eigenstates of a coupled CPB-cavity system
differs from the excitation energies to higher excited states, and
thus they form an effectively closed two-level system which may be
resonantly coupled to a collective spin oscillator.

Our objective is to store qubits for quantum computing, and in
particular to store many qubits in the same medium. To this end, we
consider the application of a magnetic field gradient to the sample
for a duration $\tau$ as described in Fig.~\ref{fig:setup}. The field
provides a linearly varying Zeeman energy shift across the sample so
that after a certain interaction time, a linearly varying spatial
phase $\exp(i k z)$, with $k=-m_0\, \Delta B \tau/L \hbar$ is encoded
in the excited state component of the individual spins at position $z$
along the sample, and the collective state $\ket{\psi_1(0)}$ is
transferred to
\begin{equation} \label{1uk}
\ket{\psi_1(k)} \equiv   \frac{1}{\sqrt{N}} \sum_q
\frac{g_q}{\bar{g}} e^{i k z_q} \Ket{\downst_1 \ldots \upst_{q}\ldots \downst_N}.
\end{equation}
Different positive and negative values of the collective spin wave
number, $k$ can be chosen and we will in the following refer to the
magnetic gradient pulse, accomplishing a specific value of the phase
gradient, as a \emph{$(k)$-pulse} applied to the system.
The key idea of the holographic quantum register is that the gradient
pulses let us successively access a number of collective excitation
modes of the same spin ensemble, such that the read-in of each new
qubit does not disturb the previously stored qubits, because only the
$k=0$ spin mode of the new qubit interacts with the cavity field.
These modes are 1-dimensional $k$-space voxels as used in magnetic
resonance imaging \cite{callaghan93}; similar
orthogonal collective atomic excitation modes are being studied for
storage of multiple modes of light
\cite{surmacz08:efficient,moiseev01:complete,hetet08:photon,zhao09:millisecond,lauro09:spectral}.

\newcommand{\scalartimes}{\times}
To identify modes which are truly independent, and which can hence be
used for storage of different qubits, we introduce the creation
operator for the $i^{\text{th}}$ spin wave mode as
\begin{equation}
  \label{eq:bkdef}
  \hat{b}^\dagger(k_i) \equiv \frac{1}{\sqrt{N}} \sum_{q=1}^N \frac{g_q}{\bar{g}} e^{i k_i z_q}\hat{\sigma}^{(q)}_+,
\end{equation}
and consider the commutator $[\hat{b}(k_i),\hat{b}^\dagger(k_j)]$ in
the fully polarized limit.
If the spins are arranged on a Bravais lattice and have equal coupling
strengths, the commutator vanishes for any pair of members of the
reciprocal lattice, so that the corresponding modes are perfectly
independent \cite{dyson56:general}.
For a general geometry, we find that $[\hat{b}(k_i),\hat{b}^\dagger(k_j)]=M(k_j-k_i)$,
where we have introduced the mode overlap
$M(\Delta k) \equiv \sum_q e^{i \Delta k\, z_q} \lvert g_q \rvert^2/N \bar{g}^2$.
For the uniformly doped $L=\lambda$ transmission line cavity
illustrated in Fig.~\ref{fig:setup}, 
$M(\Delta w\, 2 \pi/L)=\sinc(\pi \Delta w)/[1-(\Delta w/2)^2]$ 
in the continuous ($N \to \infty$) limit, so that in this case, the
mode overlap vanishes when the difference in winding number $\Delta
w\equiv \Delta k L/2 \pi$ is equal to any integer except $0$ or $\pm
2$.
It follows that choosing $k_n=3 n \times 2 \pi/L$, or even
$\{k_n\}=\{0,3,4,7,8,\ldots \}\times 2 \pi/L$, for the register modes
will ensure that $M(k_j-k_i)=0$ for all pairs of modes.
The duration of a $k=2 \pi/L$ gradient pulse is $2 \pi\hbar/m_0\Delta
B$, and to implement such a pulse in $100 \, \text{ns}$ for the system
of Fig.~\ref{fig:setup}, a field gradient of $13 \,\text{mT}/\text{m}$
is required across the sample.
A gradient of this strength allows us to sequentially address hundreds
of modes, while correspondingly stronger gradients or longer pulse
interaction times are needed to switch between any pair of modes.

\begin{figure}
  \centering
  \includegraphics[width=.9\linewidth]{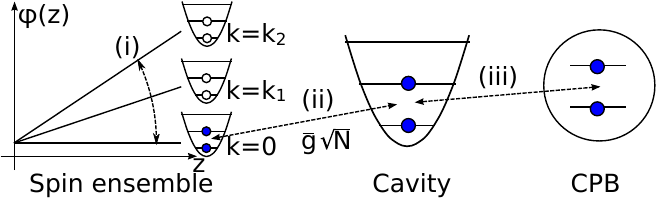}
  \caption{\label{fig:outline}  (Color online) 
    Schematic view of the quantum degrees of freedom in our
    hybrid physical system. The spatial spin modes associated with
    different $k$-values are indicated by the lines in the $\phi(z)$
    vs. $z$ diagram (left).
    (i) A $(\Delta k)$ gradient pulse interaction maps any mode $k$ to
    $k'=k+\Delta k$ [Eq.~(\ref{eq:bkdef})].  (ii) Only the $k=0$ mode
    interacts strongly with the cavity field mode [Eq.~(\ref{HamCS})]
    which, in turn, is coupled (iii) to the two level CPB system by a
    Jaynes-Cummings interaction.
  }
\end{figure}

We have now established that the $k$-modes of the spin ensemble in the
strongly polarized limit behave as a large number of independent
harmonic oscillators.
Using gradient pulses, these spatial modes can be mapped to the $k=0$
mode which is the only one that can be brought to interact strongly
with the cavity field.
When the spin ensemble is brought into resonance by adjusting the bias
field $B$, the collective coupling [Eq.~(\ref{HamCS})], swaps the
states of the $k=0$ mode and the cavity at the effective coupling rate
$\sqrt{N} \bar{g}$. We can thus selectively swap the state of the
cavity with any $k_i$ mode of the spin ensemble without disturbing any
other mode by applying first a $(-k_i)$ pulse followed by a sweep
through resonance and finally a $(k_i)$ pulse.
To extend our addressable qubit register into a full quantum
information processing system, we need to to further include a
physical qubit system which can be initialized, manipulated and read
out, and which can facilitate a suitably non-linear interaction
between the oscillator modes as outlined in Fig.~\ref{fig:outline}.
Our candidate for this qubit is the transmon CPB
\cite{koch07:charge-insensitive} which has already been integrated
experimentally with the system geometry illustrated in
Fig.~\ref{fig:setup}. Single qubit unitary operations can be achieved
by applying classical resonant fields to the CPB system, and by
swapping the states between any selected $k_i$ and the CPB degree of
freedom, and one-qubit gates as well as readout
\cite{wallraff05:approaching} can hence be applied to any qubit in our
register.
By swapping a control qubit to the CPB and
subsequently swapping a target qubit to the cavity field, the
controlled interaction between these two system effectively implements
a two-qubit gate on any pair of qubits \cite{tordrup08:holographic}.

So far, we have only discussed the coupling of the $k=0$ mode to a
quantum field in a cavity, but the individual addressability and
independence of the different $k$-modes, can also be demonstrated with
magnetic field gradients and classical resonant field pulses. A short
duration classical pulse ``tilts'' the macroscopic magnetization
vector of the sample by a small angle, generating a component of the
magnetization perpendicular to $z$ proportional to the tilt angle; the
component parallel to $z$ changes only to second order.
During a magnetic field gradient pulse, the perpendicular component of the
magnetization precesses non-uniformly across the sample, encoding the
excitation in a spin wave and reducing the total perpendicular component of
the magnetization effectively to zero.
Subsequent excitations may be stored in spin wave modes in a similar
way. Excitations stored in this way may be detected by applying a
gradient pulse that converts a particular spin wave back into a
uniform transverse magnetization. In the uniform bias field, the
uniform transverse magnetization precesses coherently and its
inductive signal may be detected using a standard pulsed magnetic
resonance spectrometer. 

Our proposal relies on several key assumptions, and we now turn to a
discussion of potential errors resulting from mechanisms bringing
their validity into question.

In order for the collective spin wave excitations to act as
independent oscillator modes, the density of excited spins must be
low. This would be adequately fulfilled at an operating temperature of
$20\,\text{mK}$ where the thermal excitation probability of spins is
$p\approx 10^{-5}$.
Even at such low temperature, a large total number $p N$ of
thermally excited spins will be present in the sample. 
The reason that we expect to be able to recognize a single
collective spin excitation on the background of, possibly, millions of
excited spins is that these spin excitations are distributed over all
spins and hence over all collective modes of the system, while the
excitation in the small number of modes that we interact with is
negligible.
For each collective
mode, the population outside the ground state will only be $\langle
\hat{b}^\dagger (k_i) \hat{b}(k_i) \rangle \approx p$. The fact that
most of the excitation resides in spectator modes, which are not
coupled to the cavity field, can, in the case of homogeneous coupling,
be argued more directly in the collective spin representation of the
system \cite{wesenberg02:mixed_collec_states_many_spins}.
Further, the independence of the modes ensures that we may actively
cool the $k_i$ modes used in the register by transferring their
excitation to the cavity field, allowing an efficient preparation of
the ground state of the register, even at finite temperatures.

The dipole-dipole interactions must be weak enough that the
spin-waves are eigenstates of the system Hamiltonian. This will not be
strictly fulfilled, and the dipole-dipole interaction will lead to
decoherence of the register.
For random doping as considered here, we expect the decoherence rate
to be on the order of the line broadening induced by random
dipole-dipole coupling
\cite{anderson51:theory,kittel53:dipol_broad_magnet_reson_lines},
while for a regular lattice of spins, the dominant decoherence
mechanism is expected to be dipole-mediated Raman scattering with
thermal spin-waves \cite{sparks60:ferromagnetic,sparks67:theory}. With
the spins distributed uniformly in a layer of $1\,\mu \text{m}$ in the
setup of Fig.~\ref{fig:setup}, the dipole-induced line
broadening would be $\sim 2\pi\times 50 \,\text{kHz}$
\cite{wesenberg04:field_insid_a_random_distr}.
Also, in a real experimental apparatus, it is difficult to make the
bias field sufficiently homogeneous that different electrons
experience the same precession, so that the collective excitation of
the ensemble remains coherent for times comparable to the intrinsic
dephasing time.

Errors due to static inhomogeneities in the external fields,
e.g.~field distortions due to the Meissner effect of the
superconducting cavity electrodes, are well known in nuclear and
electronic spin resonance studies and can be countered by applying a
classical refocusing pulse which stimulates a Hahn-echo.
The superconducting surfaces may also contribute a fluctuating
external field \cite{sendelbach08:magnetism}, the effect of which can
be minimized by not positioning spins close to electrode surfaces. As
$\sqrt{N}\bar{g}$ is independent of mode volume for constant spin
density, this can be achieved without loss of effective coupling
strength.
For the system illustrated in Fig.~\ref{fig:setup}, the refocusing
could be performed by introducing a strong microwave field at the Larmor
frequency, either through the cavity or by an alternative coupling
mechanism.
Ideally, the refocusing pulses rotate all members of the ensemble by
exactly $\pi$ about an axis perpendicular to $z$. In practice,
however, it is not possible to apply perfect $\pi$-pulses and this
strategy is expected to introduce a very large number of excitations
into the system.
It might be expected that a significant fraction of such
excitations would enter and swamp the register modes, but on closer
inspection this is not the case.
To argue this, let us for simplicity assume that the classical driving
field shares the amplitude and phase characteristics of the cavity
mode, $B_0(\vec{r})$. 
In this case, any deviation $\varepsilon_1$ ($\varepsilon_2$) of the
area of the first (second) echo pulse from $\pi$ would introduce $\sim N
\varepsilon_2^2$ excitation in the $k=0$ mode and similarly $\sim N
\varepsilon_1^2$ excitations into an ``inhomogeneous'' mode 
where the relative phase of the excited spins are given by the
inhomogeneous precession over the Hahn pulse interval.  In general, the
 ``inhomogeneous'' mode is not a $k$-mode, although in the
absence of inhomogeneities it corresponds to the $k=0$ mode. 
In the discrete spin case, the mode
overlap between any pair of modes is $\sim1/\sqrt{N}$, so that the
Hahn-echo errors will introduce $\sim \varepsilon_i^2$ excitations
into each register mode.
The echo sequence can be applied to free spin ensembles without a
cavity and CPB and the effect of the refocusing can thus be
demonstrated for classical spin waves.
The functioning of the cavity and the CPB will not be seriously
impeded, but special means may be needed to avoid excessive direct
excitation of these components of the hybrid system during the
application of the strong echo pulses.

In summary, we have proposed a quantum register capable of holding
hundreds of physical qubits in collective excitations of a spin
ensemble, and we have indicated how to perform qubit encoding, one- and
two-bit gates, and read-out in this system.
Further, we have identified mechanisms and properties of the
collective states that make the register resilient
to the effects of finite polarization and errors introduced by echo
sequences used to counter the effects of bias field inhomogeneities.

\begin{acknowledgments}
  KM is supported the EU integrated project SCALA;
  AA and JJLM by the Royal Society;
  JHW and GADB by QIPIRC (EPSRC GR/S82176/01 and GR/S15808/01);
  RJS and DIS by  NSF DMR-0653377 and Yale University; 
  JJLM by St.~John's College, Oxford; 
  and JHW by the Carlsberg Foundation.
  JHW acknowledges the hospitality of the National Research Foundation \& Ministry of Education, Singapore.
\end{acknowledgments}




\begin{thebibliography}{10}

\bibitem{brion07:quantum}
E.~Brion, K.~M{\o}lmer, and M.~Saffman, ``Quantum computing with collective
  ensembles of multilevel systems,''
  \href{http://dx.doi.org/10.1103/PhysRevLett.99.260501}{Phys. Rev. Lett. {\bf
  99}, 260501 (2007)}.

\bibitem{saffman08:scaling}
M.~Saffman and K.~M{\o}lmer, ``Scaling the neutral-atom rydberg gate quantum
  computer by collective encoding in holmium atoms,''
  \href{http://dx.doi.org/10.1103/PhysRevA.78.012336}{Phys. Rev. A {\bf 78},
  012336 (2008)}.

\bibitem{tordrup08:quantum}
K.~Tordrup and K.~M{\o}lmer, ``Quantum computing with a single molecular
  ensemble and a {C}ooper-pair box,''
  \href{http://dx.doi.org/10.1103/PhysRevA.77.020301}{Phys. Rev. A {\bf 77},
  020301(R) (2008)}.

\bibitem{tordrup08:holographic}
K.~Tordrup, A.~Negretti, and K.~M{\o}lmer, ``Holographic quantum computing,''
  \href{http://dx.doi.org/10.1103/PhysRevLett.101.040501}{Phys. Rev. Lett. {\bf
  101}, 040501 (2008)}.

\bibitem{surmacz08:efficient}
K.~Surmacz, J.~Nunn, K.~Reim, K.~C. Lee, V.~O. Lorenz, B.~Sussman, I.~A.
  Walmsley, and D.~Jaksch, ``Efficient spatially resolved multimode quantum
  memory,'' \href{http://dx.doi.org/10.1103/PhysRevA.78.033806}{Phys. Rev. A
  {\bf 78}, 033806 (2008)}.

\bibitem{koch07:charge-insensitive}
J.~Koch, T.~M. Yu, J.~Gambetta, A.~A. Houck, D.~I. Schuster, J.~Majer,
  A.~Blais, M.~H. Devoret, S.~M. Girvin, and R.~J. Schoelkopf,
  ``Charge-insensitive qubit design derived from the {C}ooper pair box,''
  \href{http://dx.doi.org/10.1103/PhysRevA.76.042319}{Phys. Rev. A {\bf 76},
  042319 (2007)}.

\bibitem{wallraff04:strong}
A.~Wallraff, D.~I. Schuster, A.~Blais, L.~Frunzio, R.-S. Huang, J.~Majer,
  S.~Kumar, S.~M. Girvin, and R.~J. Schoelkopf, ``Strong coupling of a single
  photon to a superconducting qubit using circuit quantum electrodynamics,''
  \href{http://dx.doi.org/10.1038/nature02851}{Nature {\bf 431}, 162 (2004)}.

\bibitem{schoelkopf08:wiring}
R.~J. Schoelkopf and S.~M. Girvin, ``Wiring up quantum systems,''
  \href{http://dx.doi.org/10.1038/451664a}{Nature {\bf 451}, 664 (2008)}.

\bibitem{tyryshkin03:electron}
A.~M. Tyryshkin, S.~A. Lyon, A.~V. Astashkin, and A.~M. Raitsimring, ``Electron
  spin relaxation times of phosphorus donors in silicon,''
  \href{http://dx.doi.org/10.1103/PhysRevB.68.193207}{Phys. Rev. B {\bf 68},
  193207 (2003)}.

\bibitem{morton06:electron}
J.~J.~L. Morton, A.~M. Tyryshkin, A.~Ardavan, K.~Porfyrakis, S.~A. Lyon, and
  G.~A.~D. Briggs, ``Electron spin relaxation of {N}@{C}{$_{60}$} in
  {CS}{$_2$},'' \href{http://dx.doi.org/10.1063/1.2147262}{J. Chem. Phys. {\bf
  124}, 014508 (2006)}.

\bibitem{morton07:environmental}
J.~J.~L. Morton, A.~M. Tyryshkin, A.~Ardavan, K.~Porfyrakis, S.~A. Lyon, and
  G.~A.~D. Briggs, ``Environmental effects on electron spin relaxation in
  {N}@{C}{$_{60}$},'' \href{http://dx.doi.org/10.1103/PhysRevB.76.085418}{Phys.
  Rev. B {\bf 76}, 085418 (2007)}.

\bibitem{morton08:solid-state}
J.~J.~L. Morton, A.~M. Tyryshkin, R.~M. Brown, S.~Shankar, B.~W. Lovett,
  A.~Ardavan, T.~Schenkel, E.~E. Haller, J.~W. Ager, and S.~A. Lyon,
  ``Solid-state quantum memory using the {$^{31}$}{P} nuclear spin,''
  \href{http://dx.doi.org/10.1038/nature07295}{Nature {\bf 455}, 1085 (2008)}.

\bibitem{yale:largebok}
D.~Schuster {\em et al.} In preparation.

\bibitem{chase08:collective}
B.~A. Chase and J.~M. Geremia, ``Collective processes of an ensemble of
  spin-1/2 particles,''
  \href{http://dx.doi.org/10.1103/PhysRevA.78.052101}{Phys. Rev. A {\bf 78},
  052101 (2008)}.

\bibitem{andre06:coherent}
A.~Andre, D.~DeMille, J.~M. Doyle, M.~D. Lukin, S.~E. Maxwell, P.~Rabl, R.~J.
  Schoelkopf, and P.~Zoller, ``A coherent all-electrical interface between
  polar molecules and mesoscopic superconducting resonators,''
  \href{http://dx.doi.org/10.1038/nphys386}{Nat. Phys. {\bf 2}, 636 (2006)}.

\bibitem{rabl06:hybrid}
P.~Rabl, D.~DeMille, J.~M. Doyle, M.~D. Lukin, R.~J. Schoelkopf, and P.~Zoller,
  ``Hybrid quantum processors: Molecular ensembles as quantum memory for solid
  state circuits,''
  \href{http://dx.doi.org/10.1103/PhysRevLett.97.033003}{Phys. Rev. Lett. {\bf
  97}, 033003 (2006)}.

\bibitem{imamoglu08:cavity-qed}
A.~Imamoglu, ``Cavity-{QED} based on collective magnetic dipole coupling: spin
  ensembles as hybrid two-level systems,''
  \href{http://dx.doi.org/10.1103/PhysRevLett.102.083602}{Phys. Rev. Lett. {\bf
  102}, 083602 (2009)}.

\bibitem{callaghan93}
P.~T. Callaghan, {\em Principles of nuclear magnetic resonance microscopy}.
\newblock Clarendon Press, Oxford, 1993.

\bibitem{moiseev01:complete}
S.~A. Moiseev and S.~Kr\"oll, ``Complete reconstruction of the quantum state of
  a single-photon wave packet absorbed by a doppler-broadened transition,''
  \href{http://dx.doi.org/10.1103/PhysRevLett.87.173601}{Phys. Rev. Lett. {\bf
  87}, 173601 (2001)}.

\bibitem{hetet08:photon}
G.~H\'{e}tet, M.~Hosseini, B.~M. Sparkes, D.~Oblak, P.~K. Lam, and B.~C.
  Buchler, ``Photon echoes generated by reversing magnetic field gradients in a
  rubidium vapor,'' \href{http://dx.doi.org/10.1364/OL.33.002323}{Opt. Lett.
  {\bf 33}, 2323 (2008)}.

\bibitem{zhao09:millisecond}
B.~Zhao, Y.-A. Chen, X.-H. Bao, T.~Strassel, C.-S. Chuu, X.-M. Jin,
  J.~Schmiedmayer, Z.-S. Yuan, S.~Chen, and J.-W. Pan, ``A millisecond quantum
  memory for scalable quantum networks,''
  \href{http://dx.doi.org/10.1038/nphys1153}{Nat. Phys. {\bf 5}, 95 (2009)}.

\bibitem{lauro09:spectral}
R.~Lauro, T.~Chaneli\`{e}re, and J.-L. Le~Gou\"{e}t, ``Spectral hole burning
  for stopping light,''
  \href{http://dx.doi.org/10.1103/PhysRevA.79.053801}{Phys. Rev. A {\bf 79},
  053801 (2009)}.

\bibitem{dyson56:general}
F.~J. Dyson, ``General theory of spin-wave interactions,''
  \href{http://dx.doi.org/10.1103/PhysRev.102.1217}{Phys. Rev. {\bf 102}, 1217
  (1956)}.

\bibitem{wallraff05:approaching}
A.~Wallraff, D.~I. Schuster, A.~Blais, L.~Frunzio, J.~Majer, M.~H. Devoret,
  S.~M. Girvin, and R.~J. Schoelkopf, ``Approaching unit visibility for control
  of a superconducting qubit with dispersive readout,''
  \href{http://dx.doi.org/10.1103/PhysRevLett.95.060501}{Phys. Rev. Lett. {\bf
  95}, 060501 (2005)}.

\bibitem{wesenberg02:mixed_collec_states_many_spins}
J.~Wesenberg and K.~M{\o}lmer, ``Mixed collective states of many spins,''
  \href{http://dx.doi.org/10.1103/PhysRevA.65.062304}{Phys. Rev. A {\bf 65},
  062304 (2002)}.

\bibitem{anderson51:theory}
P.~W. Anderson, ``Theory of paramagnetic resonance line breadths in diluted
  crystals,'' \href{http://dx.doi.org/10.1103/PhysRev.82.291}{Phys. Rev. {\bf
  82}, 291 (1951)}.

\bibitem{kittel53:dipol_broad_magnet_reson_lines}
C.~Kittel and E.~Abrahams, ``Dipolar broadening of magnetic resonance lines in
  magnetically diluted crystals,''
  \href{http://dx.doi.org/10.1103/PhysRev.90.238}{Phys. Rev. {\bf 90}, 238
  (1953)}.

\bibitem{sparks60:ferromagnetic}
M.~Sparks and C.~Kittel, ``Ferromagnetic relaxation mechanism for {M}{$_z$} in
  yttrium iron garnet,''
  \href{http://dx.doi.org/10.1103/PhysRevLett.4.232}{Phys. Rev. Lett. {\bf 4},
  232 (1960)}.

\bibitem{sparks67:theory}
M.~Sparks, ``Theory of three-magnon ferromagnetic relaxation frequency for low
  temperatures and small wave vectors,''
  \href{http://dx.doi.org/10.1103/PhysRev.160.364}{Phys. Rev. {\bf 160}, 364
  (1967)}.

\bibitem{wesenberg04:field_insid_a_random_distr}
J.~H. Wesenberg and K.~M{\o}lmer, ``The field inside a random distribution of
  parallel dipoles,''
  \href{http://dx.doi.org/10.1103/PhysRevLett.93.143903}{Phys. Rev. Lett. {\bf
  93}, 143903 (2004)}.

\bibitem{sendelbach08:magnetism}
S.~Sendelbach, D.~Hover, A.~Kittel, M.~Muck, J.~M. Martinis, and R.~McDermott,
  ``Magnetism in {SQUID}s at millikelvin temperatures,''
  \href{http://dx.doi.org/10.1103/PhysRevLett.100.227006}{Phys. Rev. Lett. {\bf
  100}, 227006 (2008)}.

\end{thebibliography}

\bibliographystyle{utphys-jaw}
\providecommand{\href}[2]{#2}\begingroup\raggedright\endgroup
\end{document}